\documentclass[aps,pra,reprint,groupedaddress,showpacs,longbibliography]{revtex4-1}

\usepackage{amssymb,amsmath}
\usepackage{graphicx}
\usepackage{txfonts}
\usepackage{xcolor}

\usepackage{siunitx}
\newcommand{\micron}{$\si{\micro m}$}
\newcommand{\microsec}{$\si{\micro s}$}

\begin{document}

\title{Resonant-light diffusion in a disordered atomic layer}

\author{R. Saint-Jalm$^{1}$}
\author{M. Aidelsburger$^{1,\dagger}$}
\author{J.L. Ville$^{1}$}
\author{L. Corman$^{1,\ddagger}$}
\author{ Z. Hadzibabic$^{2}$}
\author{D. Delande$^{1}$}
\author{S. Nascimbene$^{1}$}
\author{N. Cherroret$^{1}$}
\author{J. Dalibard$^{1}$}
\author{J. Beugnon$^{1}$}

\email[]{beugnon@lkb.ens.fr\hspace{0.1cm} cherroret@lkb.upmc.fr}

\altaffiliation[$^\dagger$Present address: ]{Fakult\"at f\"ur Physik, Ludwig-Maximilians-Universit\"at M\"unchen, Schellingstr. 4, 80799 Munich, Germany}
\altaffiliation[$^\ddagger$Present address: ]{Institute for Quantum Electronics, ETH Zurich, 8093 Zurich, Switzerland}

\affiliation{$^1$Laboratoire Kastler Brossel,  Coll\`ege de France, CNRS, ENS-PSL University, Sorbonne Universit\'e, 11 Place Marcelin Berthelot, 75005 Paris, France}
\affiliation{$^2$Cavendish Laboratory, University of Cambridge, J.J. Thomson Avenue, Cambridge CB3 0HE, United Kingdom}

\date{\today}
\begin{abstract}
\noindent

Light scattering in dense media is a fundamental problem of many-body physics, which is also relevant for the development of optical devices. In this work we investigate experimentally light propagation in a dense sample of randomly positioned resonant scatterers confined in a layer of sub-wavelength thickness. We locally illuminate the atomic cloud and monitor  spatially-resolved  fluorescence away from the excitation region. We show that light spreading is well described by a diffusion process, involving many scattering events in the dense regime. For light  detuned  from resonance we find evidence that the atomic layer behaves as a graded-index planar waveguide. These features are reproduced by  a simple geometrical model and numerical simulations of coupled dipoles.

\end{abstract}

\maketitle

Multiple-scattering in disordered materials is currently the focus of intense investigations in many different contexts such as electron transport in solids \cite{Akkermans07}, sound wave propagation \cite{Hu08}, matter waves in optical potentials \cite{Aspect09} and light propagation in dielectric materials \cite{vanRossum99}. Multiple light scattering is of paramount importance to understand light transport, for instance, in biological tissues, planetary atmospheres or interstellar clouds \cite{vandeHulst12}. In addition, the development of custom photonic materials allows one to engineer disordered materials in a controlled way and opens new applications in random lasing \cite{Wiersma08} or in the development of solar cells \cite{Wiersma13}.

Cold atomic gases offer a unique platform to investigate light scattering. Due to the simple atomic level structure, all photons incident on the gas are scattered without absorption. Cooling techniques can bring the atoms to a temperature where their residual motion is so small that Doppler broadening is negligible. Additionally, the gas dimensionality can be changed rather easily by shaping the trapping potential. Finally, the tunability of  the scattering cross section and the atomic density allows one to explore light scattering from the dilute to the dense regime.

In the multiple-scattering regime, light is scattered several times before exiting the material in random directions. In sufficiently dilute atomic clouds, this process is known to be well captured by a random-walk type, diffusive model \cite{Akkermans07}. This regime has been explored in several experiments, achieving for instance the observation of coherent backscattering \cite{Labeyrie99, Bidel02},  subradiant and superradiant modes \cite{Roof16, Guerin16, Araujo16}, cooperative radiation pressure force \cite{Bienaime10} or collective light scattering \cite{Balik13,Bromley16}. At higher densities $\rho$ where the product of the light wave number $k$ and the inter-atomic spacing  $ \rho^{-1/3}$  becomes of the order of unity, the situation is still poorly understood. Indeed, scatterers are no longer independent of each other and significant induced dipole-dipole interactions occur \cite{Lehmberg70,Friedberg73,Gross82,Morice95, Cherroret16}. These interactions are responsible for ``cooperative'' effects such as broadening and  shift of the resonance line \cite{Keaveney12,Javanainen14,Pellegrino14,Jennewein16,Jenkins16,Corman17}.

\begin{figure}[hbt!!!!]
\centering
\includegraphics[width=0.28\textwidth]{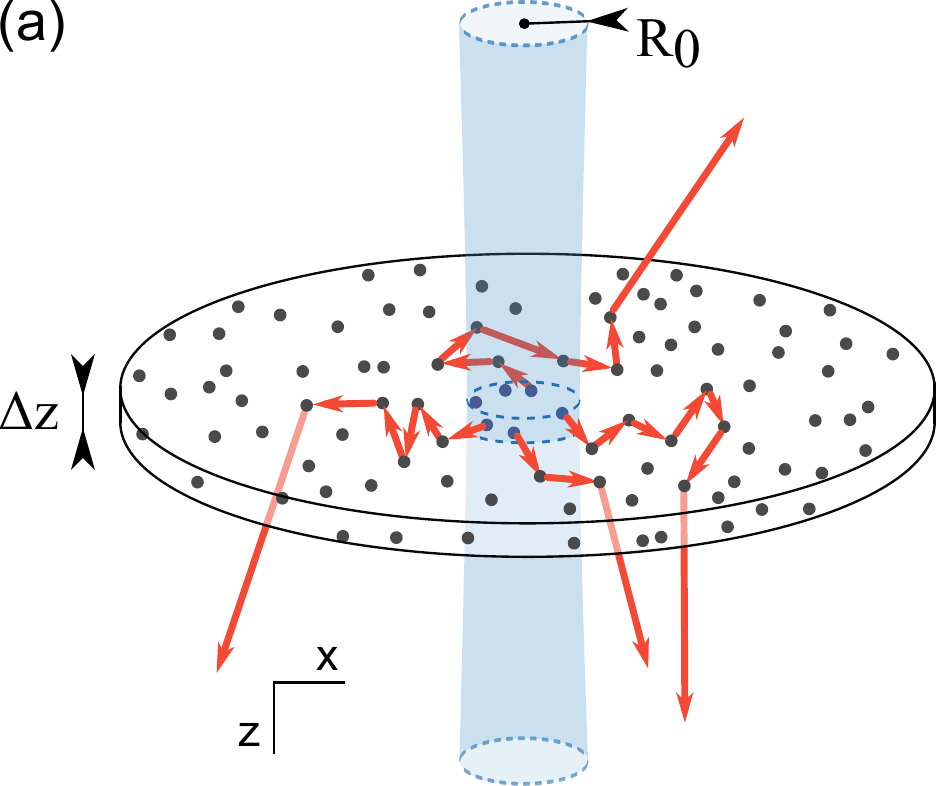}
\includegraphics[width=0.18\textwidth,trim=85 0 30 0,clip=true]{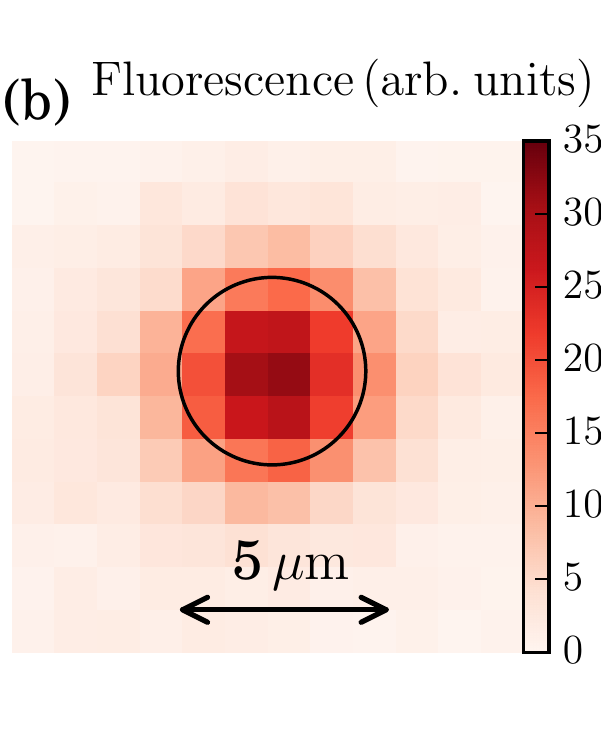}\vspace{0.7cm}
\includegraphics[width=0.48\textwidth]{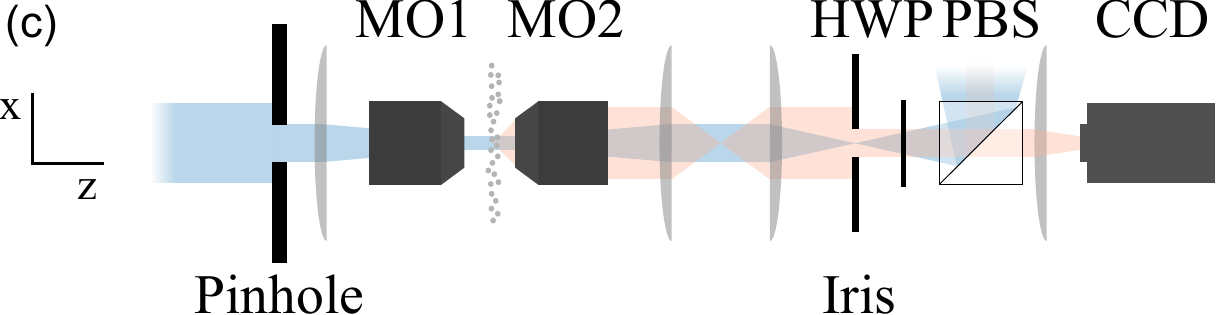}
\caption{Experimental setup. A layer of atoms is illuminated at its center with a resonant light beam of radius $R_0$. Photons initially emitted in the excitation region can be scattered several times in the atomic layer before exiting the system and being collected and imaged. (b) Typical fluorescence image at resonance and maximum density. The black circle, of diameter of $2R_0=5\,\mu$m, delineates the excitation region. Intensity is not uniform inside the excitation region partially because of finite optical resolution but also and more importantly, because of the light diffusion process itself, as described in the text. (c) Schematics of the imaging system. A pinhole is imaged on the atomic cloud thanks to a microscope objective (MO1). The atomic fluorescence is imaged on a camera (CCD) thanks to a second and identical objective (MO2) and a set of lenses. Fluorescence is spatially filtered thanks to an iris and the detected polarization is chosen by adjusting a combination of a half-wave plate (HWP) and a polarization beam splitter (PBS).}\label{fig1a}
\end{figure}


Here, we explore the phenomenon of multiple-scattering of light in a dense and large cloud of atoms.  In our ultracold sample the  motion of the atoms is negligible on the timescale of the experiment, so they act as a gas of fixed, randomly distributed point scatterers. Atoms are confined in a layer geometry in the focal plane of a high resolution imaging system, which allows us to inject light in a region with sharp boundaries and monitor its spreading away from this region (see Fig.\,\ref{fig1a}a). We observed that the fluorescence intensity, which measures the local photon escape rate, decays exponentially with the distance from the excitation zone. We show that this behavior is compatible with a two-dimensional (2D) diffusive model. For resonant light, we measured the decay length of the fluorescence signal as a function of the atomic density, all the way from the dilute to the dense regime. This decay length, or equivalently the escape radius of the photons, represents the distance the photons travel before escaping the sample. It decreases with increasing density, and then saturates in the regime of high density where photons undergo a few tens of scattering events before leaving the cloud. For detuned light and in the dense regime, the photon escape rate is significantly modified, in a way that suggests a light-guiding mechanism reminiscent of a graded-index planar waveguide. We also observed these phenomena in numerical simulations based on a model of coupled dipoles, and explained them in a semi-quantitative way by an analytical model of light guiding in an open, disordered 2D slab.

We use a dense layer of $^{87}$Rb atoms as previously described in Refs.\,\cite{Ville17,Corman17}. In the $xy$ plane, we produce a uniform disk-shaped atomic layer of radius $R=20\,\mu$m with a controllable surface density up to $\rho_{\rm 2D}=135(15)\,$\micron$^{-2}$ and a temperature $T_0=270(10)$\,nK.  Atoms are strongly confined along the vertical $z$ direction with an approximately Gaussian density distribution of r.m.s. thickness $\Delta z \approx 0.3\,$\micron$\,<\lambda_0$, with $\lambda_0=2\pi/k=0.78\,$\micron\: the resonant wavelength for the $|F=2\rangle$ to $|F'=3\rangle$ D2 transition of rubidium atoms. This corresponds to a maximum density at the center of the Gaussian profile of $\rho_{\rm 3D} k^{-3}=0.35(5)$ where $\rho_\text{3D}=\rho_\text{2D}/(\sqrt{2\pi}\Delta z)$.

We tune this density by varying the number of  atoms  in the $|F=2,m_F=-2\rangle$ hyperfine ground state, which is sensitive to our light probe. The population in this state is controlled thanks to partial transfer from the $|F=1,m_F=-1\rangle$ state in which the atoms are initially spin-polarized. As discussed in Ref. \cite{Corman17}, dipole-dipole interactions play a dominant role at the densities achieved in our setup. Because of these interactions, the thickness of the cloud could increase, depending on its density, during the light excitation. We estimate that the effective cloud thickness, in the illuminated region, could reach a maximum value of $0.4\,\mu$m at the end of the excitation.

\begin{figure}[hbt!!!!]
\centering
\includegraphics[width=0.48\textwidth]{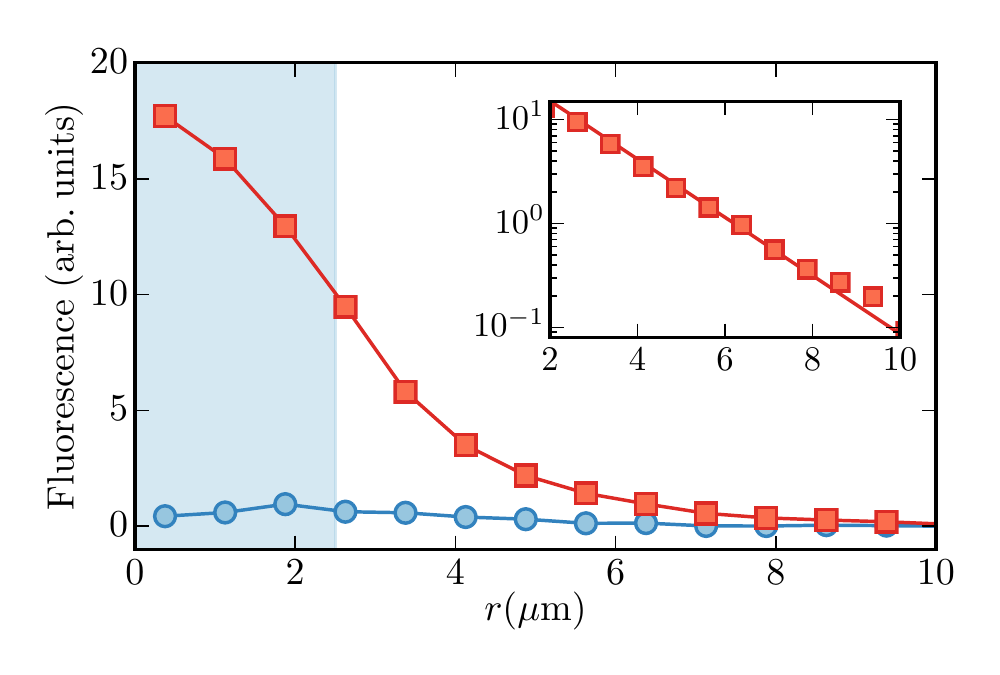}
\caption{Fluorescence decay. Binned and azimuthally averaged profile for the measured atomic fluorescence (squares). Circles show the small background signal observed without atoms. The shaded area represents the excitation region. Statistical errors bars due to photon counting are smaller than the size of the points. The solid lines are broken lines linking the points. The inset shows the same data outside the excitation region in a semilog plot to emphasize the exponential decay of the fluorescence signal. We attribute the deviation of the last two points  from the exponential fit to the contribution from stray light. Data are averaged over 100 measurements.}\label{fig1b}
\end{figure}

The atomic cloud is locally excited by light at a wavelength $\lambda_0$, propagating perpendicular to the atomic plane along the $z$ axis, as illustrated in Fig.\,\ref{fig1a}a, and linearly polarized along $x$. In the illuminated region, the intensity of the beam is on the order of 7\,$I_{\rm sat}$, where $I_{\rm sat}=1.6\,$mW\,cm$^{-2}$ is the saturation intensity for this transition (with a linewidth $\Gamma=2\pi\, \times\,$6.0\,MHz) \footnote{The intensity we used is large enough to ensure that, even if our samples have large optical depths, the  cloud is well-excited for all positions along the $z$ direction.}.  The light beam profile on the atomic cloud is given by the image of a pinhole whose diameter on the atomic cloud is $2R_0=5\,$\micron. The excitation duration is $\tau=10\,$\microsec. The atomic fluorescence is collected, spatially filtered and imaged on a CCD camera, as shown in Fig.\,\ref{fig1a}c. We detect only photons with a linear polarization perpendicular to the excitation polarization and block the residual light transmitted in the spatial mode of the incident beam. The optical resolution of our system ($\sim\!1\,$\micron) is characterized in \cite{REFSM}. Atoms outside the illuminated region are only excited by scattered light and feel a much lower intensity than in the illuminated region. Taking into account our collection and detection efficiencies, assuming that the scatterers are independent and that polarization is randomized for photons emitted from outside the excitation region, we obtain a rough estimate for the intensity seen by atoms at $1\,$\micron\: from the edge of the illuminated region of approximately $ 0.1\,I_{\rm sat}$. It is important to operate in this low saturation regime to enable comparison with the simulations described below.

We show in Fig.\,\ref{fig1a}b a typical measurement of the atomic fluorescence signal integrated over the full duration of the excitation. The circle indicates the illuminated region. Photons are detected up to several micrometers away from this region. We show in Fig.\,\ref{fig1b} a binned and azimuthally averaged profile of the fluorescence image. There is a large ratio between the atomic signal and residual stray light over the explored experimental range. Outside the illuminated region we observe an exponential decay with distance of the atomic fluorescence over almost two decades (see inset in Fig.\,\ref{fig1b}). The typical distance over which the photons travel is of a few microns.\\

\begin{figure}[htb!!!!]
\begin{flushleft}
\includegraphics[width=0.48\textwidth]{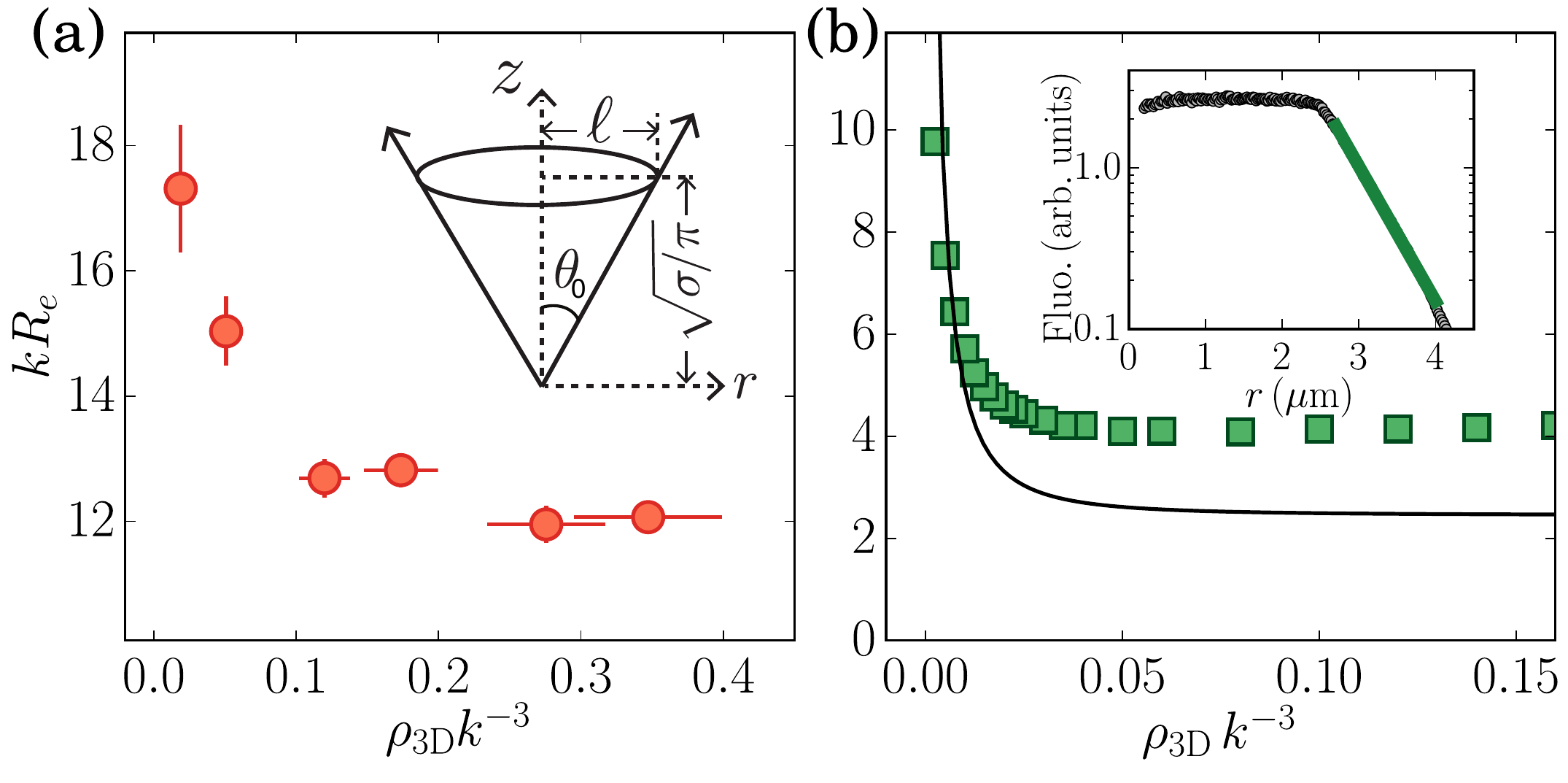}
\end{flushleft}
\caption{
(a) Measured escape radius $R_e$ for different densities at resonance ($\Delta=0$). Vertical error bars correspond to the standard deviation of the results of exponential fits to the data obtained with a bootstrap approach \cite{Efron87}. Each data set is obtained from the average of 100 measurements. Horizontal error bars represent the statistical uncertainty on the atom number. The inset shows the relevant parameters of the simple model described in the main text. (b) Escape radius $R_e$ obtained from the analytical result of Eq.\,\eqref{kxi_delta0} (solid curve) and numerical simulations of a model of classical coupled dipoles (squares) as a function of density. Statistical error bars on $R_e$ obtained from the fitting procedure of the coupled dipole simulations are smaller than the size of the points. The inset shows the fluorescence signal computed with coupled dipole simulations for $\rho_\text{3D}k^{-3}=0.12$ and averaged over $>1000$ atomic distributions. The solid line is the exponential fit outside the illuminated region.
}\label{fig2}
\end{figure}


We first focus on the experimental results for a resonant excitation. We fit the experimental fluorescence signal for various densities by $Ae^{-r/R_e}$  and obtain the escape radius $R_e$ as a function of density (see Fig.\,\ref{fig2}a). The escape radius decreases for increasing densities and reaches an approximately constant value for $\rho_\text{3D}k^{-3} > 0.1$. The measurements in the low density regime (lower values of $\rho_\text{3D}k^{-3}$) correspond to a ``single scattering event" regime where a photon typically leaves the sample after the first scattering event outside the illuminated region. In this case the escape radius is on the order of the scattering mean free path $\ell=1/(\rho_\text{3D} \sigma)$, where $\sigma$ is the light cross section and we expect $R_e \sim \ell$, in agreement with the observed decrease of $R_e$ with density.

In the opposite regime of large densities, photons are scattered several times before leaving the sample and we observe a saturation of the escape radius. It remains around $kR_e \approx 12$ while varying the density by a factor of about 3. We checked that this saturation is not due to the finite resolution of our imaging system which allows us to measure spatial structures with a size below 1\,\micron\:($kR_e \lesssim 8$). In this regime the system can be described by a diffusion model of light transport. In the steady-state regime, the light energy density $I({\bf r})$ at a point ${\bf r}$ in the sample obeys a diffusion equation with losses:
\begin{equation}
-D_0 \nabla^2 I({\bf r})=-\gamma I({\bf r})+S({\bf r}),
\end{equation}
where $D_0$ is the diffusion constant, $\gamma$ the escape rate of photons and $S({\bf r})$ the source term describing the laser excitation. We consider the situation where the source term is $S({\bf r})=S_0$ for $r<R_0$ and 0 otherwise. The solution of this equation in two dimensions is given by
\begin{equation}
I({\bf r})=\frac{S_0}{2\pi D_0} \iint_{|{\bf r'}|<R_0} K_0\left( \frac{|{\bf r-r'}|}{R_e}\right) d^2 r' \underset{r\gg R_0+R_e}{\propto}\frac{ e^{-r/R_e}}{\sqrt{r}},
\end{equation}
where $K_0$ is the modified Bessel function of the second kind of order zero and $R_e=\sqrt{D_0/\gamma}$. Outside the illuminated region, at large $r$, the function $I(r)$ decays almost exponentially \footnote{We neglect the $\sqrt{r}$ dependence which has almost no influence on the shape of the signal in the range of distance we explore in this work.} in agreement with our measurements. These results allow us to relate the measured photon escape radius $R_e$ to the diffusion constant $D_0$ and to the escape rate $\gamma$ that we cannot measure individually in a direct way. In this diffusive regime, the escape radius $R_e$ can also be related to the microscopic parameters describing the propagation of a photon in a random walk picture. Introducing $p$ the probability of escaping the cloud after a scattering event and using $\ell$ the (already defined) scattering mean free path, one gets, for large $r$,
\begin{eqnarray} 
R_e \sim \ell/\sqrt{p}.
\label{escape_rad}
\end{eqnarray}


We have developed a simple geometrical model to estimate the escape probability $p$ and to explain the main features of the curve in Fig.\,\ref{fig2}a. Consider an atom in a 2D gas emitting a photon as being placed at the center of a vertical cylinder of radius $\ell$, corresponding to the typical distance to the next scatterer. The photon will escape the cloud if its emission angle $\theta_0$ is small enough so that it would meet the next scatterer at a height larger than $\sqrt{\sigma/\pi}$. Emitted photons that escape the medium are then effectively contained within a cone of half-angle $\theta_0$ (see inset of Fig.\,\ref{fig2}a). This gives a probability $p\sim 2\times 2\pi(1-\cos\theta_0)/(4\pi)$ with $\tan\theta_0=\ell/\sqrt{\sigma/\pi}$, and in turn:
\begin{equation}
kR_e=\dfrac{k\ell}{\sqrt{1-(1+\ell^2\pi/\sigma)^{-1/2}}}\,.
\label{kxi_delta0}
\end{equation}
We display  Eq.\,\eqref{kxi_delta0} computed for the resonant scattering cross section $3 \lambda_0^2/(2\pi)$ in Fig.\,\ref{fig2}b (solid line). At large densities, $p\sim\ell^2\pi/(2\sigma)$ so that $R_e$ becomes independent of the density and much larger than the mean free path. This saturation of $R_e$ with increasing density thus stems from the compensation of two antagonistic effects: (i) the decrease of $\ell$, which tends to make the escape radius smaller and (ii) the decrease of $p$. The extension of Eq.\,\eqref{escape_rad} to low densities (large $\ell$) is consistent with $R_e \propto \ell$ and this simple model thus reproduces roughly the behavior of the ensemble of the experimental points.

Using the random walk picture and this estimate of $p$, we find  that the typical number of scattering events before a photon leaves the sample is $N_{\rm scatt}=1/p=2\sigma/(\pi \ell^2)$. Considering an atomic density of $\rho_\text{3D} k^{-3}=0.07$, in the ``plateau" of Fig.\,\ref{fig2}b but not too large so that $\ell$ could still be interpreted with the classical picture of a mean free path, we get a typical value of about $20 $ scattering events. This value justifies the diffusion model and confirms that we investigate experimentally a steady-state situation (the total duration of a photon random walk, $\approx N_{\rm scatt}\Gamma^{-1}$, is much shorter than the duration $\tau$ of the illumination pulse).

As shown in Fig.\,\ref{fig2}b, the variation of escape radius with density is also reproduced by numerical simulations based on the method of coupled dipoles. As described in Ref.\,\cite{Chomaz12}, we model our atomic system by a random ensemble of randomly positioned classical coupled dipoles (with transition $J=0\,$ to $\,J=1$) in a layer geometry with the same thickness as in the experiment. We compute the exact radiated field from these dipoles for a given excitation field taking into account simultaneously all effects related to dipole-dipole interactions and interferences. Fluorescence signals are obtained from the modulus square of each dipole and escape radii from an exponential fit to the fluorescence signal in the range $r_1=2.7\,\mu$m to $r_2=4\,\mu$m, where $r$ is the distance from the center. For each density we adapt the atom number in the simulations (from 100 up to 4000) and the number of repetitions over which the result of the simulations are averaged (from 20 to more than $10^5$).

The two models that we have developed are in  good agreement with each other. But although these predictions qualitatively reproduce  the measurements in Fig.\,\ref{fig2}a, note the difference by a factor $\approx 3$ between the scales of the two graphs. Possible reasons for this difference are discussed below.

\begin{figure}[h]
\centering
\begin{tabular}{cc}
\includegraphics[width=0.48\textwidth]{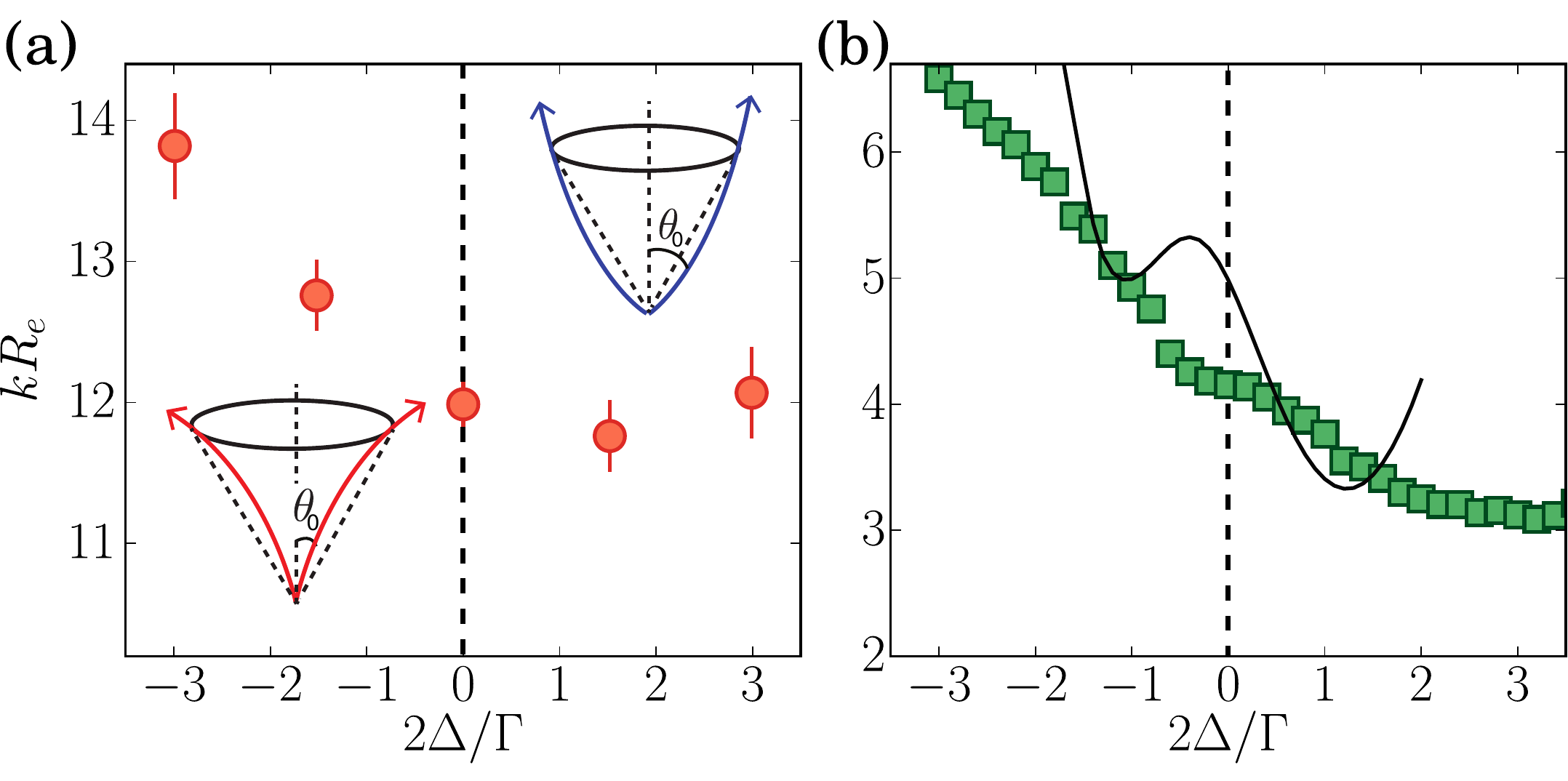}
\end{tabular}
\caption{
Escape radius versus detuning. (a) Experimental results obtained for a dense cloud with $\rho_\text{3D}k^{-3}\simeq 0.35$. The insets show the bending of the trajectories which depends on the sign of the detuning. The dashed lines in the insets correspond to the maximal emission angle a photon leaving the sample could have without considering the bending effect. Taking into account the bending effect, this maximal angle is smaller for negative detuning and  larger for positive detunings. Each point is the average result of 100 measurements. (b) Escape radius computed from a numerical model of coupled dipoles (squares), escape radius $R_e=\ell/\sqrt{1-\cos\theta_0}$ obtained by solving Eq. (\ref{eq_for_theta}) for $\theta_0$ (solid line). All theory curves are computed for the largest density we are able to handle $\rho_\text{3D}k^{-3}=0.1$. Statistical error bars obtained from the fitting procedure for the coupled dipole simulations are smaller than the size of the points.
}\label{fig4}
\end{figure}

An interesting feature of atomic systems is the possibility to change dramatically the response of the system by varying the detuning $\Delta$ of the excitation light with respect to the resonance. We report in Fig.\,\ref{fig4}a the influence of the detuning on the experimental escape radius $R_e$ for the cloud of highest density. We observe a clear asymmetry around $\Delta=0$: the escape radius is larger for negative than for positive detunings, which indicates that photons are escaping more easily the sample when $\Delta>0$.  Clearly, the results of Fig.\,\ref{fig4}a cannot be explained by the dependence of Eq.\,(\ref{kxi_delta0}) on $\Delta$, which originates only from the photon scattering cross section by a single atom and is thus symmetric with respect to $\Delta=0$.
 
We attribute the asymmetry mainly to a refractive-index gradient effect and we developed a simple model to describe this effect. We approximate the atomic slab by a continuous medium with a spatially-varying index of refraction $n(z)$ along the vertical direction and use the low-density expression for this index: $n(z)=1-[6\pi \rho_\text{3D} (z) \Delta/\Gamma]/[k^3(1+4\Delta^2/\Gamma^2)]$. It is thus either larger than one below the resonance ($\Delta<0$), or lower than one above the resonance ($\Delta>0$). In our system, the density distribution $\rho_\text{3D} (z)$ has an approximately Gaussian shape with a maximum at $z=0$ leading to a gradient of the index of refraction. For negative detuning, $n(z)$ decreases with $|z|$ from the center of the cloud. This  gives rise to a phenomenon of light guiding close to the one encountered in optical waveguides, explaining qualitatively why the escape radius gets larger. For $\Delta >0$, the opposite effect is expected, with a decrease of the escape radius.

More quantitatively, the effect of the refractive-index gradient on $R_e$ can be estimated by modifying the geometrical picture of the inset in Fig.\,\ref{fig2}a to account for the \textit{bending} of photon trajectories that escape the layer, as illustrated in the insets of Fig.\,\ref{fig4}a for $\Delta>0$ and $\Delta<0$. For negative detunings, this bending leads to a decrease of the maximum value of the emission angle $\theta_0$ for which a photon leaves the cloud. Similarly, it leads to an increase of $\theta_0$ for positive detunings. The principle of the calculation is the following. We compute  the equation for the trajectory  $z(r)$  of a photon emitted in $z=0$, $r=0$ and determine the emission angle $\theta_0$ that fulfills the ``escape condition''
\begin{equation}
z(\ell)=\sqrt{\sigma/\pi}.
\label{eq_for_theta}
\end{equation}
where $\sigma=3 \lambda_0^2/(2\pi)/(1+4\Delta^2/\Gamma^2)$. To find $z(r)$, we use the low-density expression of $n(z)$ given above, and approximate the density profile $\rho_\text{3D} (z)$ of the atomic layer by an inverted parabola. Details of the calculation and complementary data for other densities are available in \cite{REFSM}. The escape radius obtained from this model is displayed in Fig.\,\ref{fig4}b as a function of $\Delta$ and shows a significant asymmetric behavior with detuning.

A similar asymmetry is also visible in our numerical simulations of coupled dipoles (see Fig.\,\ref{fig4}b). Note that differences show up between the numerical simulations and our geometrical model. In particular, the simulations display a weaker variation with $\Delta$ than the one predicted by Eq.\,(\ref{eq_for_theta}). This effect could be attributed to light-induced dipole-dipole interactions between atoms. These interactions are known to give rise to a broadening and a blue shift of the line \cite{Corman17}, present in the simulations but not taken into account in Eq.\,(\ref{eq_for_theta}). More generally, cooperative effects could be taken into account by modifying also the mean free path and the cross section, which are fundamental parameters of our model. Calculation of these parameters in the dense regime is however a difficult task that we leave for future work. 

Since the measured escape radii are on the order of the optical wavelength, their measurement is rather challenging. We discuss two possible experimental limitations for this measurement. First, the finite resolution of the imaging system leads to an overestimate of the escape radius, but our setup benefits from a good spatial resolution, and we estimate that this correction should be at most on the order of 10\%. Second, light-induced forces caused by dipole-dipole interactions are strong when operating in the dense regime. We have thus chosen the duration of the excitation short enough to limit the atomic motion, while being long enough to probe the steady-state regime. With these parameters dipole-dipole interactions still lead to a small increase of the cloud size along the vertical axis. This increase of the cloud thickness (only in the illuminated region where the scattering rate is large) could favor the propagation of photons to a larger distance. For a pulse duration twice as long \footnote{The amplitude of the signal becomes too low for shorter pulses} we observed an increase of the escape radius $R_e$ by about 20\,\%.

Both the coupled dipole simulations and the geometrical model predict a variation of $R_e$ with density and detuning qualitatively similar to the experimental results, but with a factor $\sim\, 3 $ difference in the absolute value for $R_e$. A possible explanation for this discrepancy is the complex level structure of the rubidium atom. For instance, the averaging of the Clebsch-Gordan coefficients over all possible $\pi$ transitions relevant for our incident linearly polarized light should at least lead to a correction of the effective single-atom scattering cross section which is not taken into account in our expression of the mean free path $\ell$ \footnote{An average with equal weights on all the transition gives a cross section decreased by a factor 7/15, but it neglects any optical pumping effects during the excitation. The scattering cross section should also be modified to take into account the spectrum of the photons emitted by the illuminated region. Indeed, as the Rabi frequency of our excitation beam is on the order of $\Gamma$, we expect a modification of the fluorescence spectrum.}. Taking into account this level structure on the determination of $R_e$ requires complex simulations beyond the scope of this article \cite{Jenkins16}. An alternative approach would be to reproduce this study with atomic species like ytterbium and strontium which present well-isolated $J=0$ to $J=1$ transitions. One could also couple our method for producing thin slabs with the method to isolate an effective two-level system in a multilevel atom discussed in Refs\,\cite{skipetrov15,whiting15,whiting16} and based on the large Zeeman effect produced by an external magnetic field.

In summary, we have explored  the diffusion of light in a dense and extended sample of fixed scatterers. Experiments with cold atomic systems are usually limited to lower densities. Still, the dense regime was explored in Refs.\,\cite{Pellegrino14,Jennewein16,Jenkins16} but with a microscopic sample where light propagation cannot be investigated, and also in Ref.\,\cite{Keaveney12} but with a hot and thus Doppler-broadened cloud. Complementary studies on light transport in photonic planar waveguides have also been reported, for instance, in Ref.\,\cite{Riboli17}. However, our system reveals a unique combination of multiple-scattering, high densities and guiding effects which can be tuned rather easily. Our experiment paves the way to a deeper understanding of the propagation of light in dense samples and possibly on the role of interference-induced (localization) effects in this geometry \cite{Skipetrov16,Skipetrov14}.  Indeed, while the observation of photon localization is still elusive \cite{Skipetrov16}, presumably due to the vector nature of light detrimental to interference \cite{Skipetrov14}, the problem could be circumvented  in a 2D system illuminated by a light field linearly polarized perpendicular to the disordered plane. Our atomic system could constitute a good candidate for that objective, using the escape radius -- directly controlled by the diffusion coefficient -- as a probe for localization \cite{Cherroret08}.


\normalsize
\bibliography{fluobib}

\vspace{0.5cm}
\begin{acknowledgments}
This work is supported by DIM NanoK, ERC (Synergy UQUAM). This project has received funding from the European Union's Horizon 2020 research and innovation programme under the Marie Sk\l{}odowska-Curie grant agreement N$^\circ$ 703926. We thank S. Gigan for fruitful discussions. N.C. thanks A. Browaeys for discussions on cooperative effects, and the Agence Nationale de la Recherche (Grant No. ANR-14-CE26-0032 LOVE) for financial support. Z.H. acknowledges support from EPSRC (Grant No. EP/N011759/1) and Sorbonne Universit\'e.
\end{acknowledgments}

\end{document}